\documentstyle[12pt,aasms4]{article}

\begin{document}

\title
{The FIRST Sample of Ultraluminous Infrared Galaxies at High Redshift
I. Sample and Near-IR Morphologies}

\author{S.A. Stanford\altaffilmark{1}}

\affil{Physics Department, University of California at Davis, and the Institute of Geophysics and Planetary Physics,
Lawrence Livermore National Laboratories, Livermore, CA, 94550;
adam@igpp.ucllnl.org}

\author{Daniel Stern\altaffilmark{1,}\altaffilmark{2}}

\affil{Department of Astronomy, University of California, Berkeley, CA
94720; dan@bigz.berkeley.edu}

\author{Wil van Breugel}

\affil{Institute of Geophysics and Planetary Physics,
Lawrence Livermore National Laboratories, Livermore, CA, 94550;
wil@igpp.ucllnl.org}

\author{Carlos De Breuck}

\affil{Leiden Observatory, and Institute of Geophysics and Planetary Physics,
Lawrence Livermore National Laboratories, Livermore, CA, 94550;
debreuck@igpp.ucllnl.org}

\altaffiltext{1}{Visiting Astronomer, Infrared Telescope Facility,
University of Hawaii, NASA}

\altaffiltext{2}{current address: Jet Propulsion Laboratory,
California Institute of Technology, Mail Stop 169-327, Pasadena, CA 91109}

\begin{abstract}

We present a new sample of distant ultraluminous infrared galaxies.
The sample was selected from a positional cross--correlation of the
$IRAS$ Faint Source Catalog with the FIRST database.  Objects from
this set were selected for spectroscopy by virtue of following the
well-known star-forming galaxy correlation between 1.4 GHz and 60
$\mu$m flux, and by being optically faint on the POSS.  Optical
identification and spectroscopy were obtained for 108 targets at the
Lick Observatory 3~m telescope.  Most objects show spectra typical of
starburst galaxies, and do not show the high ionization lines of
active galactic nuclei.  The redshift distribution covers $0.1 < z <
0.9$, with 13 objects at $z > 0.5$ and an average redshift of $\bar{z}
= 0.31$.  $K$-band images were obtained at the IRTF, Lick, and Keck
observatories in sub-arcsec seeing of all optically identified
targets.  About 2/3 of the objects appear to be interacting galaxies,
while the other 1/3 appear to be normal.  Nearly all the identified
objects have far-IR luminosities greater than $10^{11} L_\odot$, and
$\sim$25 \% have $L_{FIR} > 10^{12} L_\odot$.

\end{abstract}


\section{Introduction}

Observations by the Infrared Astronomy Satellite ($IRAS$) led to the
discovery of a class of galaxies with enormous far-IR luminosities.
Subsequent observations over a large range of wavelengths have shown
that these objects, called ULIG for ultraluminous infrared galaxies,
have 1) bolometric luminosities and space densities comparable to
those of optical quasars (Sanders et al.\ 1988); 2) a broad range in
host galaxy spectral type, including starburst galaxies, Seyfert I and
II, radio galaxies, and quasars; 3) morphologies often suggestive of
recent interactions or merging (Carico et al.\ 1990; Leech et al.\
1994; Rigopoulou et al.\ 1999); and 4) large amounts of molecular gas
concentrated in small ($<$1 kpc) central regions (e.g.\ Scoville et
al.\ 1989; Solomon et al.\ 1997).  Understanding the nature of the
prime energy source in ULIG has proven difficult (e.g.\ Smith,
Lonsdale, \& Lonsdale 1998).  Many of the observed characteristics
indicate that very strong starbursts could be the culprit.
Alternatively, an active galactic nucleus (AGN) may power the ULIG
(e.g.\ Lonsdale, Smith, \& Lonsdale 1993).  The very high luminosities
suggest an evolutionary connection between ULIG and quasars, wherein a
dust-enshrouded central massive black hole is gradually revealed as
the appearance of the object changes from ULIG to quasar (Sanders et
al.\ 1988).

Much effort has been expended in trying to determine the primary
source of energy---starbursts or AGN---driving the large FIR
luminosities.  The recent studies using ISO indicate that the vast
majority of the power comes from starbursts in $\sim80\%$ of the
observed systems (Genzel et al.\ 1998; Lutz et al.\ 1998).  Rigopoulou
et al.\ (1999) present the results of an expanded version of the mid-IR
spectroscopic survey first reported by Genzel et al.\ (1998).  Using
ISO to observe 62 ULIG at $z < 0.3$, they measured the line to
continuum ratio of the 7.7 $\mu$m polycyclic aromatic hydrocarbon
(PAH) feature to differentiate between starburst and AGN as the
dominant source of the large FIR luminosity.  PAH features have been
shown to be strong in starburst galaxies and weak in AGN (Moorwood
1986; Roche et al.\ 1991).  Rigopoulou et al.\ confirmed the results
of Genzel et al.\ (1998), and also found, based on near-IR imaging,
that approximately 2/3 of their sample have double nuclei and nearly
all the objects show signs of interactions.  For a recent review of
ULIG see Sanders \& Mirabel (1996).

ULIG are also of great interest for studies of early star formation in
the building of galaxies.  Recent sub-mm observations suggest that
objects similar to ULIG may contain a significant fraction of the star
formation at high redshifts (e.g.\ Lilly et al.\ 1999).  But so far
most studies have found ULIG only in the nearby universe.  Sanders et
al.\ (1988) initially studied a group of 10 objects at $z < 0.1$.
Previously published systematic surveys have found objects mostly at
$z < 0.4$ (Leech et al.\ 1994; Clements et al.\ 1996a, 1996b).  A few
high redshifts objects have been found, all of which turn out to
contain hidden AGN.  These include FSC~15307+3252 at $z=0.926$ (Cutri
et al.\ 1994) and FSC~10214+4724 at $z=2.286$ (Rowan-Robinson et al.\
1991).  The former object was found to exhibit a highly polarized
continuum, indicating the presence of a buried quasar (Hines et al.\
1995) while the latter was found to be lensed (Eisenhardt et al.\
1996) and also shows signs of containing a hidden AGN (Lawrence et
al.\ 1993; Elston et al.\ 1994; Goodrich et al.\ 1996).  Further
progress in this field has been hampered by the lack of identified
ULIG at moderately high redshifts.

No new deep far-IR survey will become available prior to the launch of
{\it SIRTF}, which will be capable of studying ULIG in detail at high
redshifts.  So, the $IRAS$ database remains the primary source of
targets for finding high redshift ULIG.  Radio observations provide a
relatively unbiased method for extracting FIR galaxies from the $IRAS$
Faint Source Catalog (FSC; Moshir et al.\ 1992) because radio
continuum emission is relatively unaffected by extinction in dense gas
and dust.  Such FIR/radio samples are ideal for detailed
investigations of the complex relationships between the interstellar
media, starbursts, and possible AGN in ULIG.  For example, a sample of
radio-loud objects was constructed by cross-correlating the $IRAS$ FSC
with the Texas 365 MHz radio catalog (TXFS; Dey \& van Breugel 1990).
Subsequent optical identifications and spectroscopy showed that the
TXFS objects tend to be distant AGN.  So a radio-quiet sample,
extracted from the FSC, should be an excellent means of finding ULIG
without AGN---i.e.\ powered by starbursts---at interesting
cosmological distances.  In this paper, we report on such a sample: we
describe the sample selection process and discuss the near-IR imaging.
We defer a detailed analysis of the radio properties and optical
spectroscopy to future papers.

\section{The FIRST/FSC Sample}

We have used two large area surveys in the radio and far-IR, which we
briefly describe here, to select ULIG candidates.  In the radio, we
have used the FIRST (Faint Images of the Radio Sky at Twenty cm;
Becker, White, \& Helfand 1995).  Using the VLA, this project is
surveying $\pi$ steradians down to a 5$\sigma$ limit of 1 mJy with 5
arcsec resolution and subarcsec positional accuracy.  One of the
problems with finding distant ULIG using $IRAS$ is that there are many
faint galaxies visible in a deep optical image within the relatively
large error ellipse of an FIR source.  The high resolution and good
positional information of FIRST offer an excellent means of choosing
the best of the many optical candidates on which to spend valuable
large telescope time getting redshifts.  We used the second version of
the catalog (released 1995 October 16), which samples 2925 degrees$^2$
in two regions of sky in the North ($7^h20^m < $ RA(J2000) $<
17^h20^m$, $22\fdg2 < $ Dec(J2000) $< 42\fdg5$) and South ($21^h20^m <
$ RA(J2000) $< 3^h20^m$, $-2\fdg5 < $ Dec(J2000) $< 1\fdg6$) Galactic
Caps.  In the far-IR we have used the $IRAS$ FSC (Moshir et al.\ 1992)
which resulted from the Faint Source Survey (FSS).  Relative to the
$IRAS$ Point Source Catalog, the FSS achieved better sensitivity by
point-source filtering the detector data streams and then coadding
those data before finding sources.  At 60~$\mu$m (the band used for
defining our candidates), the FSC covers the sky at $|b|
\gtrsim 10\fdg0$ and has a reliability (integrated over all
signal-to-noise ratios) of $\gtrsim 94\%$.  The limiting 60~$\mu$m flux
density of the FSC is approximately 0.2 Jy, where the signal-to-noise
ratio (SNR) is $\sim$5.  The FSS also resulted in the Faint Source
Reject file which contains extracted sources not in the FSC with an
SNR above 3.0.  We used the FSR, in addition to the FSC, with part of
FIRST to increase the number of targets in the fall sky.

\begin{figure}
\epsscale{1.0}
\caption{A plot of F$_\nu$ at 1.4 GHz against F$_\nu$ at 60$\mu$m
showing all matches of the positional cross-correlation of FIRST with
the $IRAS$ FSC (the open triangles), along with several comparison
samples taken from the literature.  The well-known radio--FIR
correlation for star-forming galaxies extends to the faintest flux
limits probed by the FF objects. }
\label{f20v60}
\plotone{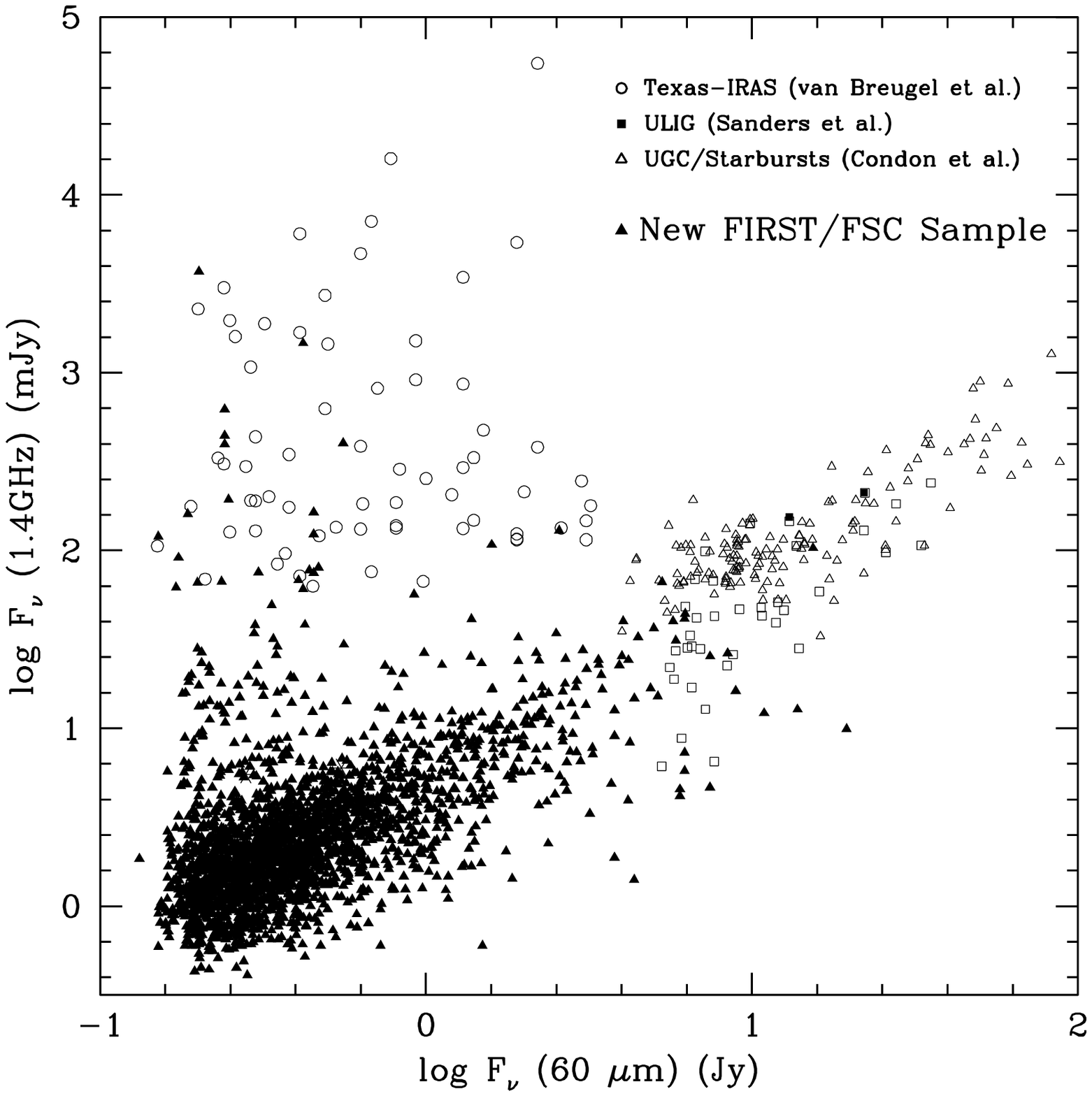}
\end{figure}

The $IRAS$ FSC was positionally cross-correlated with the second
version of the FIRST catalog, with the requirements that an FSC source
must have a real 60 $\mu$m detection ($f_{qual} \geq 1$) and that it
be within 60 arcsec of the FIRST source.  The 60 $\mu$m band was
chosen because it is more reliable than the 100 $\mu$m band and
samples close to the wavelength peak of the ULIG power.  The resulting
FIRST-FSC (FF) catalog contains 2328 matches.  To increase the
available objects in the fall sky, we also performed a positional
match of the FSR with the South Galactic Cap portion of FIRST, which
yielded an additional 176 matches.  The 20 cm and 60 $\mu$m flux
densities for this sample of 2504 sources are plotted in
Figure~\ref{f20v60}.  The majority of the FF sources fall along the
well-known radio-FIR correlation (Condon et al.\ 1991), extending from
nearby starburst galaxies to much fainter FIR/radio flux levels.  The
surface density of such objects is approximately 1 degree$^{-2}$ down
to the $\sim 5 \sigma$ limits of 1 mJy at 20~cm and $\sim$0.2 Jy at 60
$\mu$m.

We generated optical finding charts using the Digitized Sky Surveys,
available from the Space Telescope Science Institute, for all 2504
matches. The radio source position and the FSC error ellipse were
overlaid on these charts.  Visual inspection of these finding charts
was carried out to select optically faint targets for further study,
with the expectation that such targets would be distant ULIG.
Approximately 150 targets, which will carry the designation FF along
with the usual coordinate naming scheme, were selected in this manner.
A strict cutoff in optical magnitude was not employed, and we make no
attempt to construct a sample which has a well-defined limiting
magnitude in the optical.  In practice, the magnitude of the targets
selected for optical imaging and spectroscopy depended on the
observing conditions, i.e.\ some targets which are not visually faint
on the DSS image were observed during cloudy conditions.  While the
FIRST and FSC catalogs do have well-defined flux limits, our sample
was not constructed in order to be complete to a chosen flux level in
either the radio nor the far-IR bands.  The main goal of the survey is
simply to find high-redshift ULIG.  It is worth noting that our target
list would include objects with observed characteristics in the radio,
optical, and far-IR similar to those of FSC10214+4724 (which itself
lies outside of the FIRST area that we used and so cannot fall into
our catalog).  We have not found any ULIG at redshifts as great as
that of FSC10214+4724 in the $\sim$3000 degree$^2$ surveyed.
  
\section{Observations}

During several runs from March 1996 to April 1999, the Kast
spectrograph (Miller \& Stone 1994) at the Shane 3~m telescope of Lick
Observatory was used to obtain optical images and spectroscopy of the
candidate ULIG from our FF catalog.  The observing procedure typically
consisted of taking two 300~s images in the $r_S$ band, identifying
the optical counterpart of the FF source in these data, and
immediately following up with slit spectroscopy of the optical object.
Because the resolution and positional accuracy of FIRST are high, it
was usually clear which optical object coincided with the radio
source.  The FWHM of the seeing in the images was usually in the range
1.5--2.0 arcsec.  Standard stars were observed in imaging mode when
conditions were photometric.  However, because much of the data were
obtained during non-photometric conditions, $r_S$ magnitudes will not
be presented here for the sample.  Unless the source morphology
demanded a particular value, the position angle of the slit was set to
the parallactic angle.  The object was dithered along the slit by
$\sim 10$ arcsec between two exposures to aid in fringe subtraction.
Optical spectra of 1200-6000s duration were obtained of the optical
source using the 300 line mm$^{-1}$ grating in the red-side
spectrograph, which provides $\sim$4.6 \AA~pixel$^{-1}$ resolution
from 5070--10590 \AA, and a 452/3306 grism in the blue-side
spectrograph which provides $\sim$2.5 \AA~ pixel$^{-1}$ resolution
from 3000--5900 \AA.  The slit width was set at 2 arcsec.  The images
and spectra were reduced using standard techniques.

Near-infrared images were obtained of the targets for which redshifts
had been determined in order to better ascertain the morphologies of
the galaxies.  $K'$ images were obtained for nearly all identified
targets with NSFCAM at the IRTF 3~m telescope in 1998 August and 1999
February.  Additional observations of 3 targets were obtained in
service mode in September 1999.  NSFCAM was used in its 0.3 arcsec
pixel$^{-1}$ mode which provides a 77$\times$77 arcsec field.  Typical
total exposure times per object were 960s; more distant objects were
observed for twice this period.  Conditions were photometric with
seeing averaging 0.9 arcsec.  Observations of standard stars from the
Persson et al.\ (1998) list were obtained and used to calibrate the
images onto the California Institute of Technology (CIT) system, which
is defined in Elias et al.\ (1982).  The data were reduced using
standard techniques.

Five targets were observed in the $K$ band using Gemini (McLean et
al.\ 1993) at the Shane 3~m telescope on 1998 October 7.  Gemini has
0.68 arcsec pixels which give it a 174 arcsec field.  Objects were
observed for 1080~s each in photometric conditions with seeing of
$\sim$1.2 arcsec.  The data were reduced using standard techniques and
calibrated onto the CIT system using observations of UKIRT faint
source standards (Casali \& Hawarden 1992).

Two distant targets were imaged at the Keck I telescope with NIRC
(Matthews \& Soifer 1994) in 1998 April.  FF1106+3201 was observed in
the $K$ band for 16 minutes and FF1614+3234 was observed for 32
minutes in the $K_s$ band.  Both objects were observed in clear
conditions with $\sim$0.5 arcsec seeing.  These data were reduced
using standard techniques and calibrated onto the CIT system using
observations of UKIRT faint source standards (Casali \& Hawarden
1992).

\section{Results}

\subsection{Optical}

We attempted spectroscopic observations of approximately 150
IRAS/FIRST candidates, of which 116 yielded redshift information.  The
108 with infrared imaging are listed in Table 1; the 8 sources with
redshifts but lacking infrared images are not considered further.  The
sources which did not provide useful spectra were usually observed in
poor conditions; the reasons for their lack of redshifts were not
because of having intrinsically challenging spectra.  The object names
in Table 1 are based on the FIRST radio position.  The source in the
FIRST catalog would have the name given by the object's coordinates
shown in our Table 1, in the format FIRST J{\it hhmmss.s+ddmmss} where
the coordinates are truncated, not rounded.  In the $IRAS$ FSC, the
FIR source name is different from that implied by our FF name, so we
have included the FSC source name as a column in Table 1.  The Z
designation in the FSC name means that the FIR source is from the FSR
catalog.

\begin{figure}
\epsscale{1.0}
\caption{Spectra of four objects from Table 1 which cover the range in 
redshift, spectral type, and signal to noise of the sample.}
\label{spex}
\plottwo{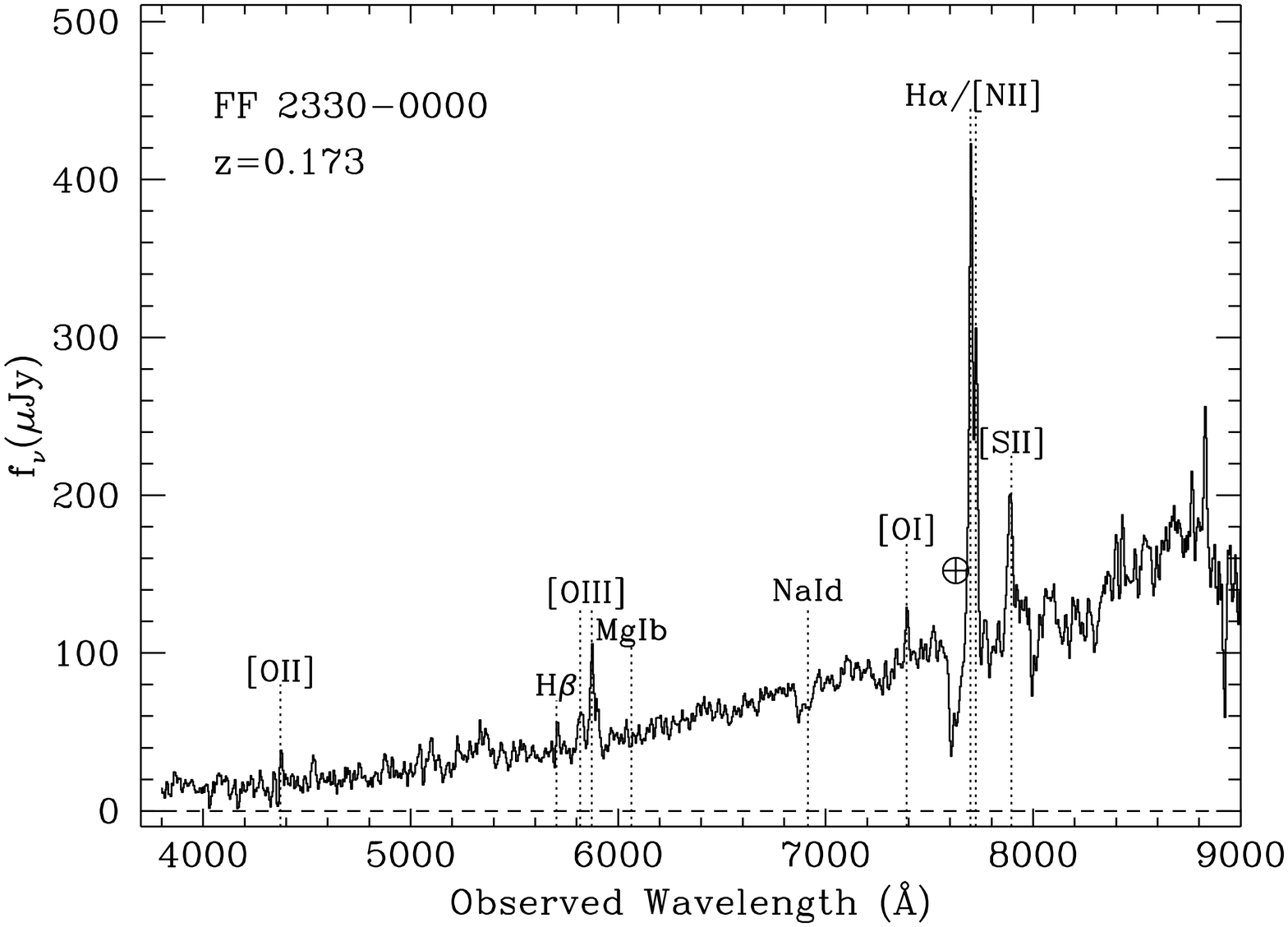}{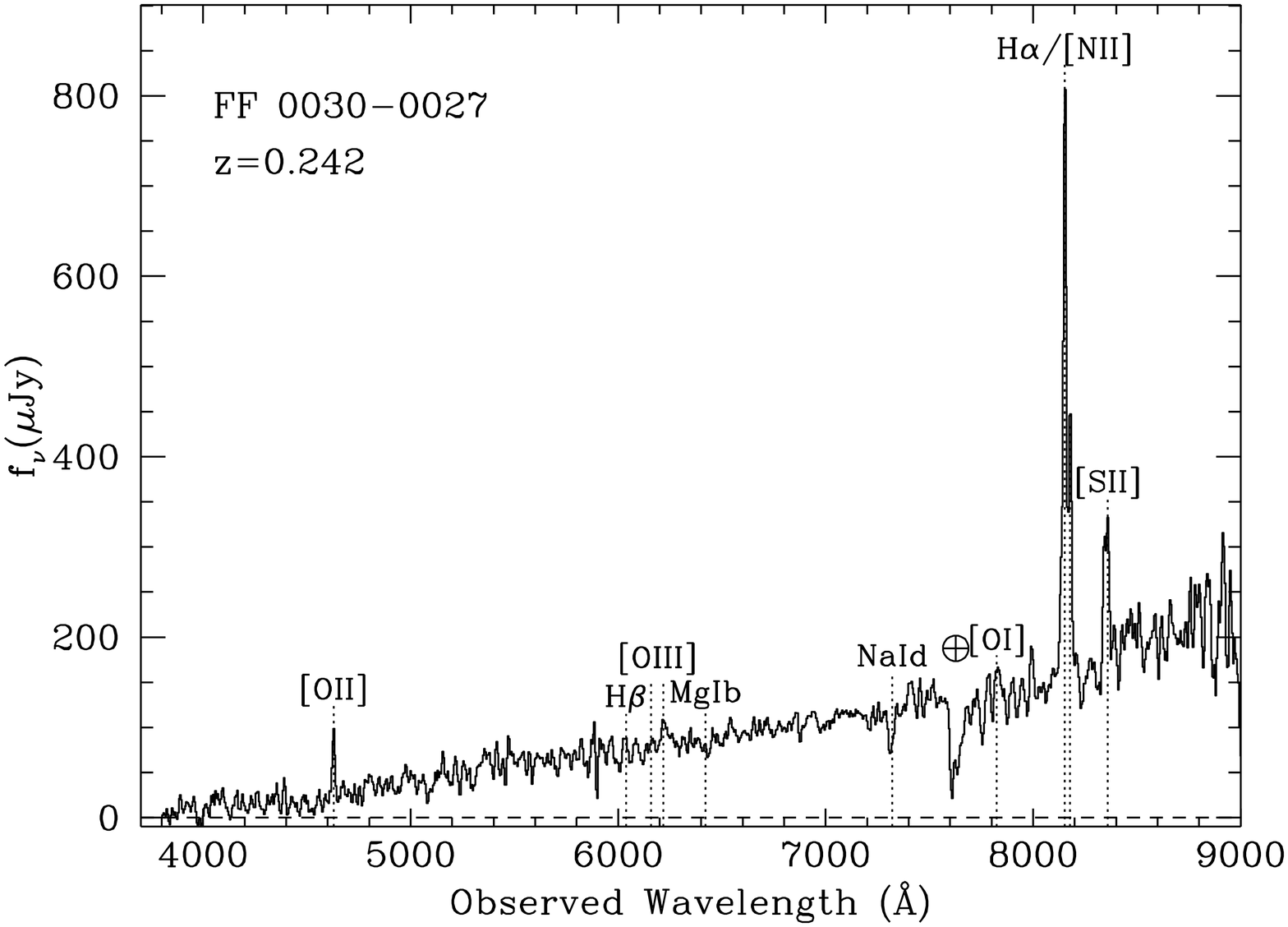}
\plottwo{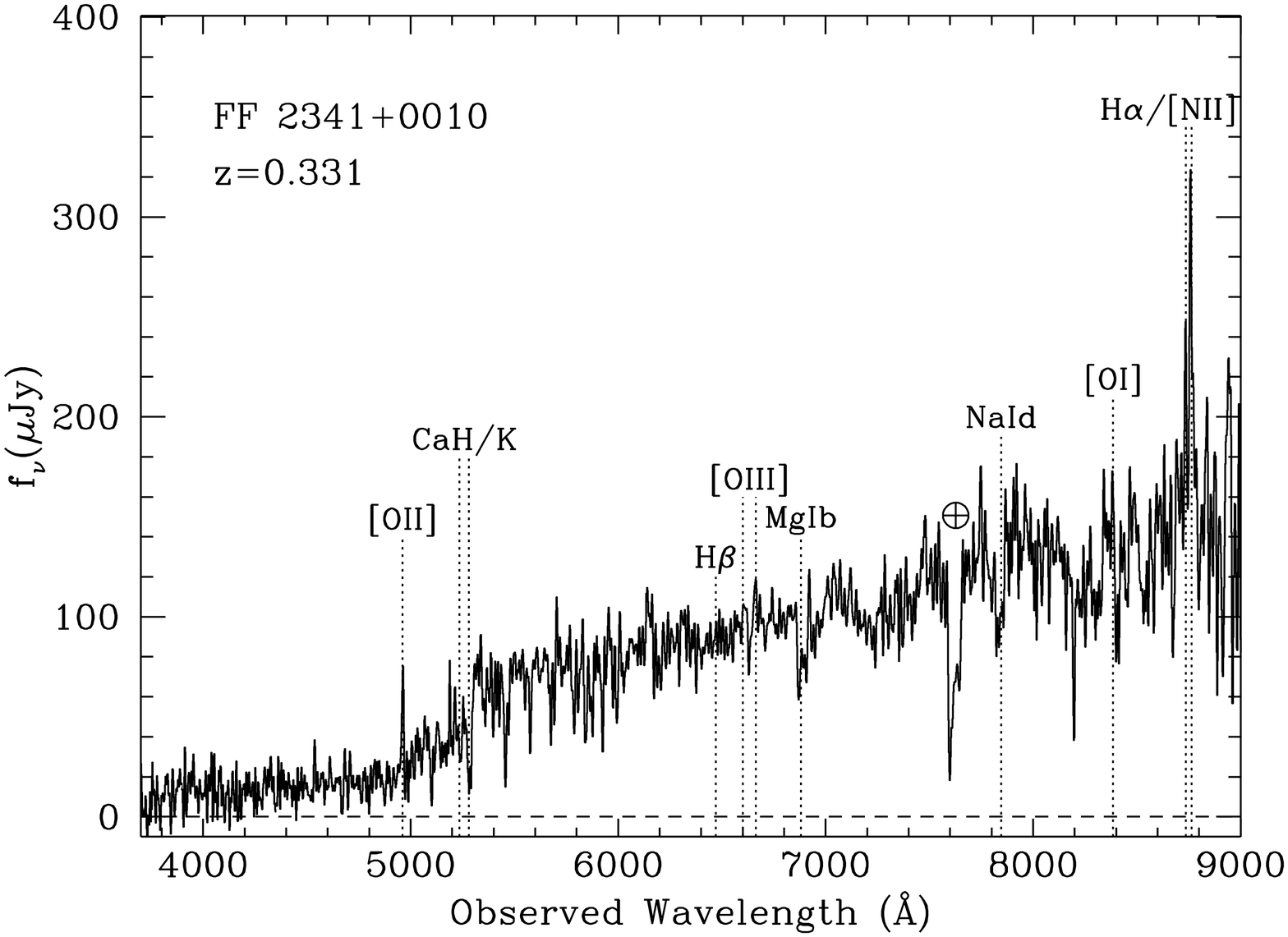}{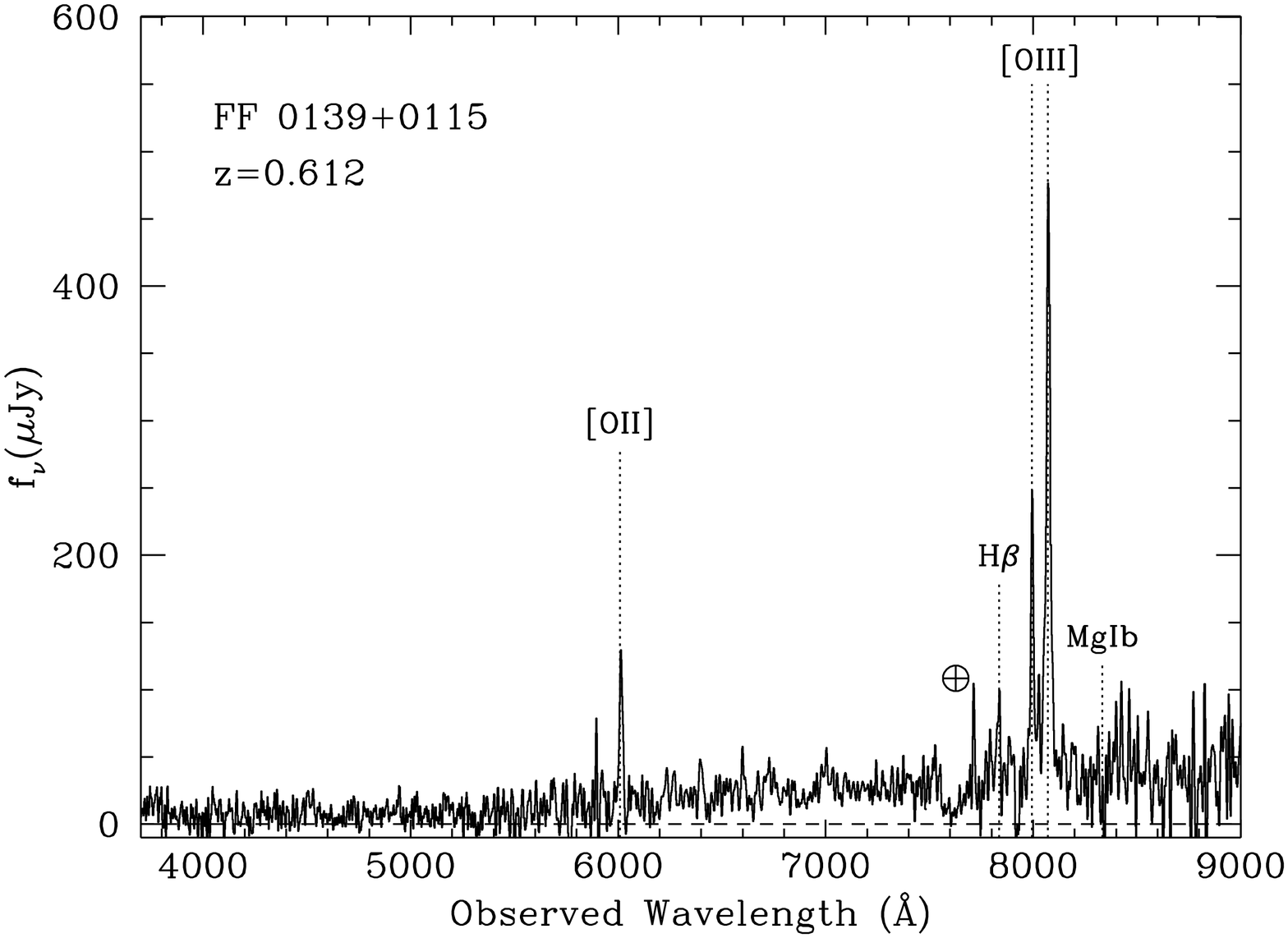}
\epsscale{1.0}
\end{figure}

The typical resolution of the spectroscopy was $\approx 15$
\AA\ (FWHM) at $\lambda > 5500$ \AA, implying typical uncertainties of
$\lesssim 0.002$ in redshift. Redshifts were determined from the spectra
after identifying probable emission lines and continuum features.  In
practice the features most often used were the [O II]$\lambda$3727, [O
III]$\lambda$4959,5007, and H$\alpha$ lines, and the D4000 break.  The
vast majority of the spectra have the emission lines characteristic of
star formation; very few show any signs, such as high ionization
lines, of an AGN.  Four sample spectra, covering a range in redshift,
signal to noise, and spectral type, are shown in Figure~\ref{spex}.  A more
detailed analysis of the optical spectra is deferred to a later paper.

\subsection{Near Infrared}

The $K'$ images are displayed for each object, along with the optical
finding chart from the DSS, in Figure~\ref{optir}.  Photometry of the FF objects
was obtained from the $K'$ images.  In Table 1, the $K$ magnitudes
within 3 arcsec diameter apertures, centered on the peak of the
near-IR emission, are given for each object.  The 5 $\sigma$ detection 
limit in most of the images is $K \sim 19$ so the limiting factor in
the uncertainty of the photometry is not the signal to noise, since
most objects have magnitudes some 3--4 magnitudes brighter than the
detection limit, but rather systematics in the zeropoint.  We estimate 
that the uncertainty in the zeropoint is $0.03$ mag.  For most objects the 3
arcsec diameter aperture contains $\sim 70\%$ of the total light.  The
morphologies of the objects tend to show signs of galaxy interactions,
including tidal tails, multiple nuclei, and disturbed outer envelopes.
Approximately 2/3 of the sample show such features, while 1/3 of the
sample appear to be normal galaxies.  A brief description of the
near-IR morphology for each FF is included in Table 1.

\subsection{Radio}

One of the major advantages of using FIRST in our survey is the high
accuracy of its positional information.  The coordinates listed in
Table 1 are those of the radio source as given by the FIRST catalog,
which has an absolute astrometric uncertainty of $\sim$1 arcsec.  The
20~cm VLA images of all objects listed in Table 1 were extracted from
the FIRST database.  The radio morphologies were classified by visual
inspection of these cutout images, and by consulting the deconvolved
sizes listed in the FIRST catalog.  The 20~cm morphological
information is given for each FF source in Table 1.  The 20~cm flux
densities listed in Table 1 have typical uncertainties of 10\% at the
2 mJy level.

\subsection{Far Infrared}

\begin{figure}
\caption{A plot of the log P$_{1.4 GHz}$ against log L$_{FIR}$
showing our new FF sample as solid triangles, along with several
comparison samples taken from the literature.  Three well-known ULIG
are marked by name.  Radio-loud quasars and BL Lacs are included at
the most luminous end of the track traced by the Texas-FSC sources.
The luminosities for F10214 have been corrected for gravitational
lensing.}
\label{p20vlfir}
\plotone{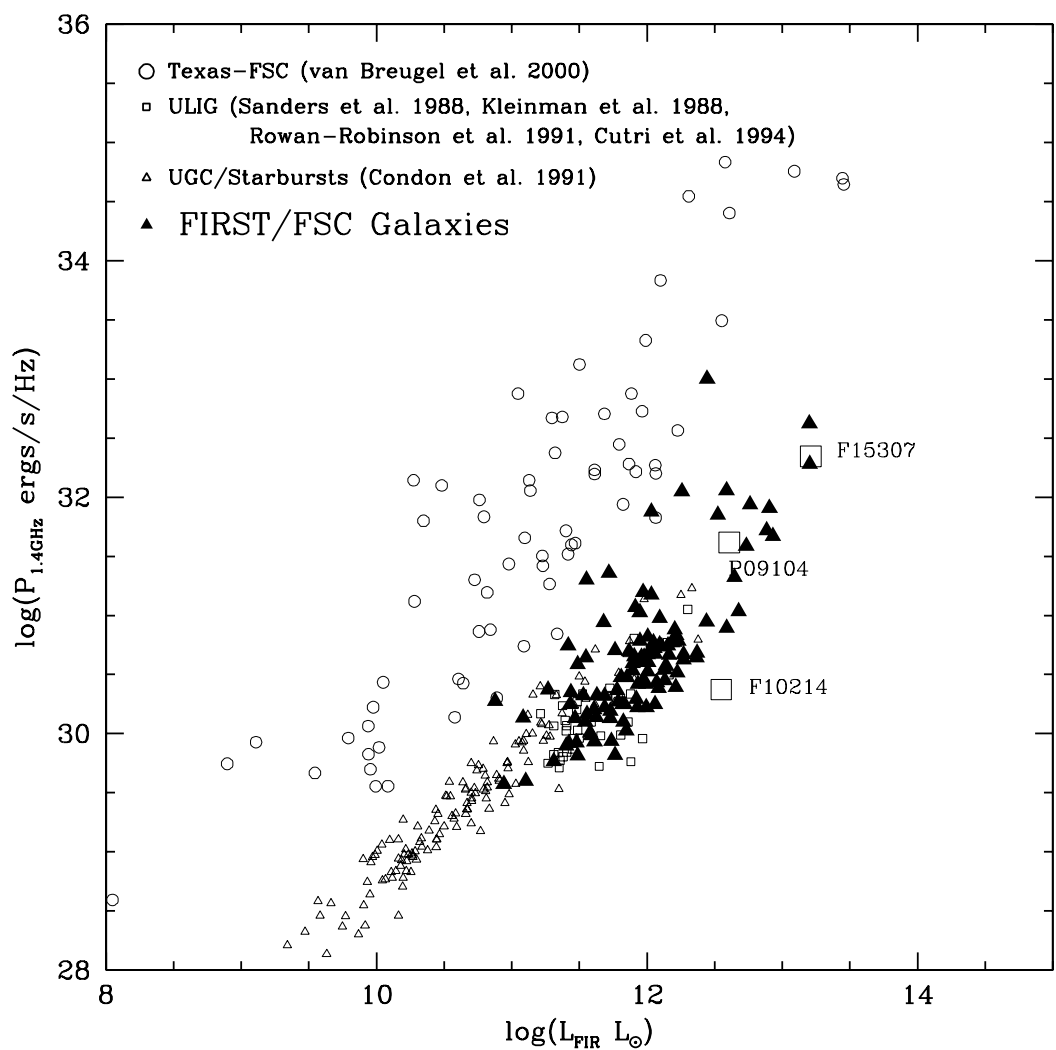}
\end{figure}

Improved $IRAS$ flux densities were obtained for all objects in Table
1 with the ADDSCAN utility at IPAC.  In addition to the 60~$\mu$m band
used to construct our FF catalog, data at 12~$\mu$m, 25~$\mu$m, and
100~$\mu$m was searched for detections.  Almost none of the objects in
Table 1 were detected at either of the shorter two wavelengths, so no
information is included from these wavebands in Table 1.  Many
detections were obtained in the 100~$\mu$m data; these are included
where available in Table 1, and one $\sigma$ upper limits are
indicated in parentheses.  The uncertainty in the typical 60~$\mu$m
measurement in the sample is $\sim 10\%$, and $\sim 15\%$ in the
100~$\mu$m band where detected.  The 60~$\mu$m and 100~$\mu$m flux densities were
used to calculate the far-infrared luminosity, as defined by Sanders
\& Mirabel (1996): L(40--500 $\mu$m) $= 4 \pi ~ D^2_L ~ C ~ F_{FIR}
[L_\odot]$, where $D^2_L$ is the luminosity distance in Mpc, $F_{FIR}
= 1.26 \times 10^{-14} (2.58 \times f_{60} + f_{100}) [W m^{-2}]$, and
$C = 1.6$.  Throughout this paper we use H$_0 = 65$ km s$^{-1}$
Mpc$^{-1}$ and $\Omega_M = 0.3$ with $\Lambda = 0$.  When 100~$\mu$m
detections could not be obtained with ADDSCAN, the 1$\sigma$ limiting
flux densities were used in the calculation of the L$_{FIR}$.  The
L$_{FIR}$ are given in Table 1, and are plotted in
Figure~\ref{p20vlfir}.  The uncertainty in the L$_{FIR}$ is dominated
by a combination of the typical $\sim 10-15\%$ flux measurement
errors and the $\sim 15\%$ uncertainty in the scaling factor C
which accounts for the extrapolated flux longward of the 100 $\mu$m
band.

\section{Discussion}

The reliability of the optical identification with the radio source
for the objects in Table 1 is very high.  The optical/radio source
association with the FSC far-infrared source is less certain, because
of the relatively large positional uncertainty of the $IRAS$
detections.  But there are at least four reasons to believe that the
identified optical/radio sources are indeed the FSC sources as well.
First, in all cases, the FIRST position is within twice the 1 $\sigma$
error ellipse of the FSC source.  Second, the optical spectra show
emission lines typical of star-forming galaxies, as expected for most
far-IR luminous objects.  Third, in the cases where both 60 and 100
$\mu$m detections were obtained, the $IRAS$ flux ratios are typical of
FIR luminous galaxies (Soifer et al.\ 1987).  Finally, in nearly all
cases the FIR/radio flux ratio lies on the well-established
correlation as seen in Figure~\ref{f20v60}.

There are at least two possible ways that the wrong association is
being made.  First, a galaxy optically brighter than the identified
FF source could lie just outside the $IRAS$ error ellipse and still be the
source of the IRAS detection.  But in our sample there are no such
galaxies which are also detected by FIRST, as would be expected if such
objects were the true source of the FIR emission.  Second, a faint
radio source could be missed by the FIRST survey that would coincide with
the $FSC$ source.  Using the rms given in the FIRST catalog for each
object, lower limits to the resulting $S_{60\mu{\rm m}}/S_{20cm}$
ratios for objects beneath the $FIRST$ detection limits are found to
be $\sim$500-700, far greater than the standard ratio for star-forming
galaxies.  Finally its worth noting that on average one expects to
find only 0.035 FIRST sources within the typical IRAS error ellipse
area, so the probability of a random radio source being associated with
an FSC source is low.  In our sample of 108 identified FF objects,
approximately 4 could be due to radio sources unassociated with the
far-IR source.

Although we defer the scientific analysis of the new sample to future
contributions, we will briefly compare some basic global properties of
our FF sample to those of other similarly large samples of high
redshift ULIG which have been previously published, e.g.\ Leech et
al.\ (1994) and Clements et al.\ (1996).  Our new sample has ULIG at
higher average redshift ($\bar{z} \sim 0.3$) than that of Leech et
al.\ ($\bar{z} \sim 0.17$) and that of Clements et al.\ ($\bar{z} \sim
0.21$).  As for the interaction rate, on which there has been
some disagreement in the literature (Sanders \& Mirabel 1996), our
result that 2/3 of the sample shows signs of galaxy interactions 
is in agreement with Leech et al.\ but not Clements et al., who found
that $\sim 90\%$ of their sample were interacting systems.  Finally,
the rate of AGN-type optical spectra in our sample, which is only
$10\%$, is somewhat less than the $\sim 25\%$ found by Clements et al.  We
do see a higher incidence of AGN-type spectra at the highest L$_{FIR}$
as has been noted previously by several studies; for a review of this
topic see Sanders
\& Mirabel (1996).

\section{Summary}

\begin{figure}
\caption{A redshift histogram of the identified FF targets.}
\label{zhist}
\plotone{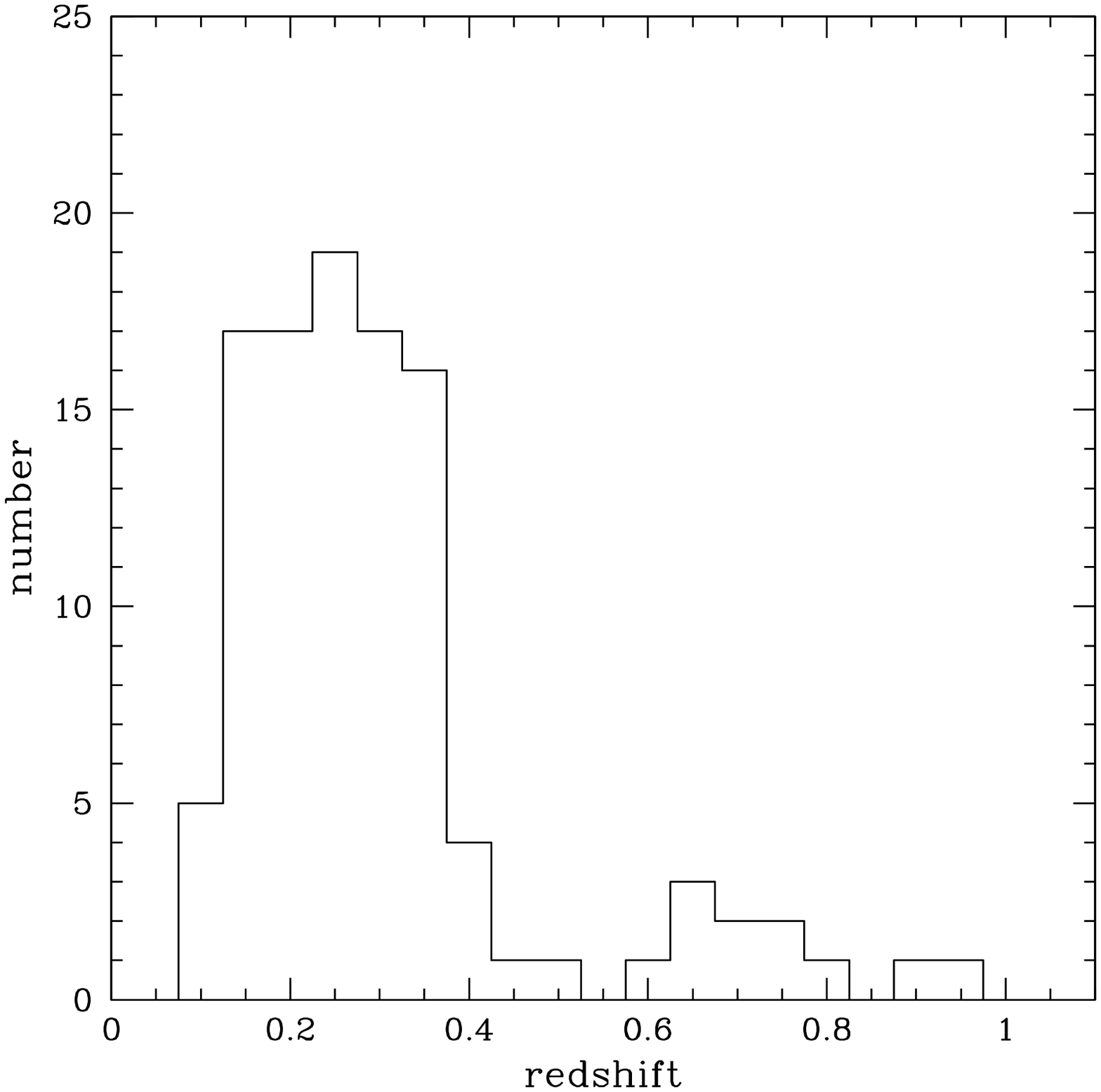}
\end{figure}

We have constructed a new survey of ULIG using a match of the $IRAS$
FSC with the second version of the FIRST catalog, which covered nearly
3000 degrees$^2$.  By choosing for further study only optically faint
matches from the DSS which also fall on the radio-FIR flux
correlation, we have attempted to find high redshift ULIG which are
powered primarily by starbursts.  Optical images and spectra were
obtained of 108 such targets, which were found to lie in the redshift
range $0.1 < z < 0.9$; a redshift histogram is shown in Figure~\ref{zhist}.
Nearly all of these targets have $L_{FIR}$ greater than $10^{11}
L_\odot$, and have a higher average redshift of $\bar{z}=0.31$ than in other
recent ULIG surveys.  Near-IR imaging shows that while more than the
majority of objects show clear signs of galaxy interactions, nearly
1/3 appear to be normal at arcsec resolution in the $K$ band.  With
this sample, we intend to examine the nature of ULIG evolution in
future contributions.

\acknowledgments

The authors thank the staff of Lick Observatory for their help in
obtaining the optical data, and Bill Vacca and Dave Griep at the IRTF
for their assistance in obtaining the near-IR data, and for conducting
servicing observing for three of the sample objects.  We have made use
of the online facilities provided by IPAC.  We also thank Bob Becker
for providing us with the FIRST catalog and for help in using its
contents.  Finally we thank Rob Kennicutt for a speedy referee
report.  The Digitized Sky Surveys were produced at the Space
Telescope Science Institute under U.S. Government grant NAG-W-2166.
The images of these surveys are based on photographic data obtained
using the Oschin Schmidt Telescope on Palomar Mountain and the UK
Schmidt Telescope.  The plates were processed into the compressed
digital form with permission of these institutions.  The work by SAS,
WvB, and CDB at IGPP/LLNL was performed under the auspices of the
U.S.\ Department of Energy under contract W-7405-ENG-48 to the
University of California.  The work by DS was supported by IGPP grants
98-AP017 and 99-AP026.

\begin{figure}
\caption{(a-bb) For each of the FF objects in Table 1 we show on the left the DSS
finding chart centered on the FIRST radio position, which is marked 
by a small circle of diameter 5 arcsec, with the FSC 95\% confidence
limit error ellipse superposed.  The astrometric uncertainty between
the optical and radio reference frames is $\sim$1 arcsec.  On the
right in each panel is a closeup view of the $K$-band image of the
identified source.  North is up and East to the left in all panels. }
\label{optir}
\end{figure}

\begin{references}

\reference{B95}Becker, R.\, White, R.\, \& Helfand, D.\ 1995, \apj, 450,
559

\reference{C90}Carico, D., Graham, J. R., Matthews, K., Wilson, T. D., Soifer, B. T.,
Neugebauer, G., \& Sanders, D. B. 1990, \apj, 349, L39

\reference{C92}Casali, M.\ \& Hawarden, T.\ 1992, JCMT-UKIRT Newsletter, 4, 33

\reference{C96a}Clements, D., Sutherland, W. J., Sauders, W., Efstathiou, G. P.,
McMahon, R. G., Maddox, S., Lawrence, A., \& Rowan-Robinson, M. 1996a, \mnras, 279, 459

\reference{C96b}Clements, D., Sutherland, W. J., McMahon, R. G., \& Saunders, W. 1996b, \mnras, 279, 477

\reference{C91}Condon, J. J., Anderson, M. L., \& Helou, G. 1991, \apj, 378, 65

\reference{C94}Cutri, R. M., Huchra, J. P., Low, F. J., Brown, R. L., \& vanden Bout,
P. A. 1994, \apj, 424, L65

\reference{D90}Dey, A.\ \& van Breugel, W.\ 1990, IAU Colloquium 124, p.\ 309  

\reference{E96}Eisenhardt, P. R., Armus, L, Hogg, D. W., Soifer, B. T., Neugebauer,
G., \& Werner, M. 1996, \apj, 461, 72

\reference{E82}Elias, J.H., Frogel, J.A., Matthews, K., \& Neugebauer, 
G.\ 1982, AJ, 87, 1029

\reference{E94}Elston, R., McCarthy, P. J., Eisenhardt, P., Dickinson, M., Spinrad,
H., Januzzi, B. T., \& Maloney, P. 1994, \aj, 107, 910

\reference{G98}Genzel, R.\ et al.\ 1998, \apj, 498, 579

\reference{G96}Goodrich, R. W., Miller, J. S., Martel, A., Cohen, M. H., Tran, H. D.,
Ogle, P. M., \& Vermeulen, R. C. 1996, \apj, 456, L9

\reference{H95}Hines, D. C., Schmidt, G. D., Smith, P. S., Cutri, R. M., \& Low,
F. J. 1995, \apj, 450, L1

\reference{K88}Kleinmann, S.G., Hamilton, D., Keel, W.C., Wynn-Williams, C.G., Eales, 
S.A., Becklin, E.E., \& Kuntz, K.D.\ 1988, \apj, 328, 161

\reference{L93}Lawrence, A., et al.\ 1993, \mnras, 260, 281

\reference{L94}Leech, K.J., Rowan-Robinson, M., Lawrence, A., \& Hughes, J. D. 1994, \mnras, 267, 253

\reference{L99}Lilly, S.J., Eales, S.A., Gear, W.K.P., Hammer, F., Le Fevre, O.,
Crampton, D., Bond, J.R., \& Dunne, L.\ 1999, \apj, 518, 641

\reference{L93}Lonsdale, C., Smith, H., \& Lonsdale, C.\ 1993, \apj,
405, L9

\reference{L98}Lutz, D., Spoon, H. W. W., Rigopoulou, D., Moorwood, A. F. M., \&
Genzel, R. 1998, \apj, 505, 103

\reference{M94}Matthews, K.\ \& Soifer, B. T.\ 1994, Experimental Astronomy, 3, 77

\reference{M93}McLean, I.S.\ et al.\ SPIE, 1946, 513

\reference{MS94}Miller, J.S.\ \& Stone, R.P.S.\ 1994, Lick Observatory Technical Reports \# 66

\reference{M86}Moorwood, A.F.M.\ 1986, A\&A, 166, 4

\reference{M92}Moshir, M., Kopman, G., \& Conrow, T. 1992, $IRAS$ Faint Source Survey, 
Explanatory Supplement, Version 2

\reference{P98}Persson, S. E., Murphy, D. C., Krzeminski, W., Roth, M., \& Rieke,
M. J. 1998, \aj, 116

\reference{R99}Rigopoulou, D., Spoon, H.W.W., Genzel, R., Lutz, D., Moorwood, A.F.M.,
Tran, Q.D.\ 1999, \aj, 118, 2625

\reference{R91}Roche, P.F., Aitken, D.K., Smith, C.H., \& Ward, M.J.\ 1991, \mnras,
248, 606

\reference{RR91}Rowan-Robinson, M.\ et al.\ 1991, Nature, 351, 719

\reference{S88}Sanders, D.B., Soifer, B. T., Elias, J. H., Madore, B. F., Matthews,
K., Neugebauer, G., \& Scoville, N. Z. 1988, \apj, 325, 74

\reference{S96}Sanders, D. B. \& Mirabel, I. F. 1996, ARAA, 34, 749
 
\reference{S89}Scoville, N. Z., Sanders, D. B., Sargent, A. I., Soifer, B. T., \&
Tinney, C. G. 1989, \apj, 345, L25

\reference{S98}Smith, H.E., Lonsdale, C.J., \& Lonsdale, C.J.\ 1998,
ApJ, 492, 137 

\reference{S87}Soifer, B.T., Sanders, D. B., Madore, B. F., Neugebauer, G.,
Danielson, G. E., Elias, J. H., Lonsdale, C. J., \& Rice, W. L. 1987, \apj, 320, 238

\reference{S97}Solomon, P.M., Downes, D., Radford, S.J.E., \& Barrett, J.W.\ 1997,
\apj, 487, 144

\reference{S98}Surace, J., Sanders, D. B., Vacca, W. D., Veilleux, S., \& Mazzarella, J.
M. 1998, \apj, 492, 116

\reference{V00}van Breugel, W.\ 2000, ``Proceedings on Ultraluminous Galaxies:
Monsters or Babies'', (Kluwer: Germany), eds.\ L.\ Tacconi and D.\
Lutz, in press


\end{references}
\end{document}